\documentclass[aps,prl,amsmath,amssymb,superscriptaddress,showpacs,twocolumn,floatfix]{revtex4}
\usepackage{amsfonts, relsize, textcomp}
\usepackage{graphics}
\usepackage{graphicx}
\usepackage{hyperref}

\begin{document}
\title{Bilayer graphene with parallel magnetic field and twisting: Phases and phase transitions in a highly tunable Dirac system}
\author{Bitan Roy}
\affiliation{ National High Magnetic Field Laboratory, Florida State University, Florida 32306, USA}
\affiliation{ Condensed Matter Theory Center, Department of Physics, University of Maryland, College Park, MD 20742, USA}

\author{Kun Yang}
\affiliation{ National High Magnetic Field Laboratory and Department of Physics, Florida State University, Florida 32306, USA}

\date{\today}

\begin{abstract}
The effective theory for bilayer graphene (BLG), subject to parallel/in-plane magnetic fields, is derived. With a sizable magnetic field the trigonal warping becomes irrelevant, and one ends up with two Dirac points in the vicinity of each valley in the low-energy limit, similar to the twisted BLG. Combining twisting and parallel field thus gives rise to a Dirac system with tunable Fermi velocity and cutoff. If the interactions are sufficiently strong, several fully gapped states can be realized in these systems, in addition to the ones in a pristine setup. Transformations of the order parameters under various symmetry operations are analyzed. The quantum critical behavior of various phase transitions driven by the twisting and the magnetic field is reported. The effects of an additional perpendicular fields, and possible ways to realize the new massive phases is highlighted.
\end{abstract}


\maketitle

\vspace{6pt}

Carbon-based layered materials opened a new frontier in condensed matter physics, where the underlying honeycomb lattice stands responsible for Dirac or Dirac-like fermionic excitations\cite{review-graphene}. Two well studied members of this new class of materials are single-layer and bilayer graphene. Despite the fascinating potential of exhibiting various ordered phases\cite{graphene-interaction-1,graphene-interaction-2} and related quantum critical phenomena\cite{sO3-AF, supercondcriticality}, the Dirac points of monolayer graphene are remarkably stable due to a large quasi-particle Fermi velocity ($v_F \sim 10^6 m/s$); thus far, ordered phases have only been realized in the presence of (perpendicular) magnetic fields\cite{goerbig}. In this regard, bilayer BLG appears to be propitious, and has already exhibited phenomena strongly suggestive of spontaneous symmetry breaking\cite{BLG-experiment1, BLG-experiment2, BLG-experiment3, BLG-experiment4, BLG-experiment5}, possibly realizing a subset of all the possible ordered states available for the fermion to condense into\cite{roy-classification}. But the role of the mesoscopic environment, such as gate configuration, substrate etc., on the nature of the ordered states still lacks a clear understanding\cite{throckmorton-vafek}. As a result, realization of several interesting ordered states and tuning this system across (quantum) phase transitions are still among future prospects. We here propose that BLG, when immersed in parallel magnetic fields and twisted\cite{twistdefinition}, yields a unique opportunity to explore some of these interesting possibilities.

Pristine BLG is well described by a two-band model, with quadratic touching of the valence and the conduction bands. Subject to in-plane magnetic fields, each parabolic band touching (PBT) in BLG splits into \emph{two} Dirac cones\cite{yakovenko}. A similar scenario also arises when BLG is twisted, if the twisting is commensurate\cite{castro-twisted-1, sanjose-guinea-gonzalez}. However, such twofold splitting competes with the trigonal warping (TW)\cite{TWdefinition}, which, on the other hand, breaks each PBT into \emph{four} Dirac cones\cite{falko-Mccann, lemonik-lifshitz, nilsson}. We here show that when a sufficiently strong in-plane field is applied, one ends up with only two Dirac cones; this happens within accessible magnetic field strength when the field is applied along certain optimal directions (see Fig.~1). More importantly, the field/twisting controls the Fermi velocity of the resultant Dirac points, and thus the (effective) interaction strength; this allows us to tune the system across various transitions between the weak-coupling phase (where interactions are irrelevant\cite{graphene-interaction-1}) to various ordered phases. Moreover, these setups admit additional fully gapped phases that do not have any analogy in either single-layer graphene or pristine BLG. When a perpendicular magnetic field is present\cite{twisted-LL-1,twisted-LL-2}, even a richer set of new ordered phases may be realized.

Recently there has been a surge of theoretical\cite{castro-twisted-1, sanjose-guinea-gonzalez, twisted-LL-2, mele-twisted, macdonald-twisted, suarez, gonzalez} and experimental\cite{twisted-LL-1, Li-fermivelocity, twisted-experiment-1, twisted-experiment-2, Helin} activities in twisted BLG (mostly focusing on single-particle physics thus far), while BLG with in-plane field has attracted little attention thus far\cite{yakovenko}. One of the motivations of the present work is to point out their similarity, and more importantly the fact that they are complementary to each other: Twisting gives rise to a larger effects and does not couple to electron spin, but is discrete. In-plane field has a weaker effect and couples to electron spin as well, but can be tuned continuously, a virtue important for exploring critical phenomena. We show that by combining these two we have a highly tunable Dirac system ideal for exploring various phases and phase transitions.

In terms of the low-energy degrees of freedom of AB-stacked BLG, we define a 4-component spinor $\Psi^\top(\vec{k})=\big[ v_1(\vec{K}+\vec{k})$, $v_2(\vec{K}+\vec{k})$, $v_1(-\vec{K}+\vec{k})$, $v_2(-\vec{K}+\vec{k}) \big]$. For now, we suppress the fermion's spin degrees of freedom and $v_1(v_2)$ is the fermionic annihilation operator on the B sublattice in layer $1(2)$\cite{AsitesBLG}. In this basis the non-interacting Hamiltonian, in the vicinity of two valleys at $\pm \vec{K}$, reads as\cite{czetkovic-throckmorton-vafek}
\begin{equation}\label{HamilBLGoriginal}
H^{0}_{BL}= \gamma_2 \frac{k^2_x-k^2_y}{2 m}- \gamma_1 \frac{ 2 k_x k_y}{2m}+ v_2 i \gamma_0 \left( \gamma_1 k_x + \gamma_2 k_y \right),
\end{equation}
where $m=t_\perp/(2 v^2_F)$, and $v_F=3t/(2 a)$. $t$ and $a$ are respectively intralayer hopping amplitude, and lattice spacing. The interlayer nearest-neighbor hopping is $t_\perp \approx t/10$, and $v_2 \sim v_F/30$ describes the TW.\cite{band-parameters-1} Five mutually anticommuting $\gamma$-matrices are $\gamma_0=\sigma_0 \otimes \sigma_3$, $\gamma_1=\sigma_3 \otimes \sigma_2$, $\gamma_2=\sigma_0 \otimes \sigma_1$, $\gamma_3=\sigma_1 \otimes \sigma_2$, and $\gamma_5=\sigma_2 \otimes \sigma_2$, where $\vec{\sigma}$ are the standard two-dimensional Pauli matrices and $\sigma_0$ is the unity matrix. In the absence of the TW ($v_2=0$), $H^{0}_{BL}$ describes PBTs at $\pm \vec{K}$.\cite{vafek-RG} The TW splits each of the PBTs into four Dirac cones, out of which one is isotropic, while the remaining three are anisotropic, connected by $120^o$ rotations about the isotropic one.\cite{ lemonik-lifshitz, nilsson}

Subject to an in-plane magnetic field $\vec{B}=B \left( \cos \theta, \sin \theta, 0\right)$, the above Hamiltonian conforms to\cite{supplementary}
\begin{eqnarray}\label{effectiveMag}
H[B] =H^{0}_{BL} +  \gamma_2 \; \bigg( \frac{k^2_{Bx}-k^2_{By}}{2 m} \bigg) - \gamma_1 \left( \frac{ 2 k_{Bx} k_{By}}{2m}\right),
\end{eqnarray}
where $\vec{k}_{B}= (d/2) (\hat{z}\times\vec{B})$, and $d \sim 3.5 \mathring{A}$ is the interlayer separation.\cite{band-parameters-2} Here $\theta$ is measured w.r.t. the momentum axis $k_x$, shown in Fig.~1(left). If $v_2=0$, $H[B]$ describes isotropic massless Dirac fermionic excitations in the vicinity of four points $\pm \vec{K} \pm (k_{Bx} \hat{k}_x+ k_{By} \hat{k}_y)$. The effect of the in-plane magnetic field (or twisting) is, therefore, qualitatively similar to the nematic order, which split the two PBTs to four Dirac cones by spontaneously breaking lattice rotation symmetry\cite{vafek-kun}. The in-plane field and twisting, which explicitly breaks lattice rotation symmetry, can thus be viewed as a field coupled to the nematic order parameter (and therefore forces a non-zero value on it). When $v_2$ is finite, one of the anisotropic Dirac cones and the isotropic one get pushed towards each other by the in-plane field, while the remaining two move apart from each other. If the in-plane magnetic field is applied along the line connecting the isotropic and one of the anisotropic Dirac points, i.e., $\theta=(\pi, 5\pi,9 \pi )/6$, the TW becomes irrelevant for $B$ $>B_{C}$ $\sim 25$ T, with the currently estimated strength of various band parameters\cite{band-parameters-1, band-parameters-2}, shown in Fig. 1(right). For the rest of the discussion, we set $v_2=0$.
\begin{figure}[htb]
\includegraphics[width=4.25cm,height=4.0cm]{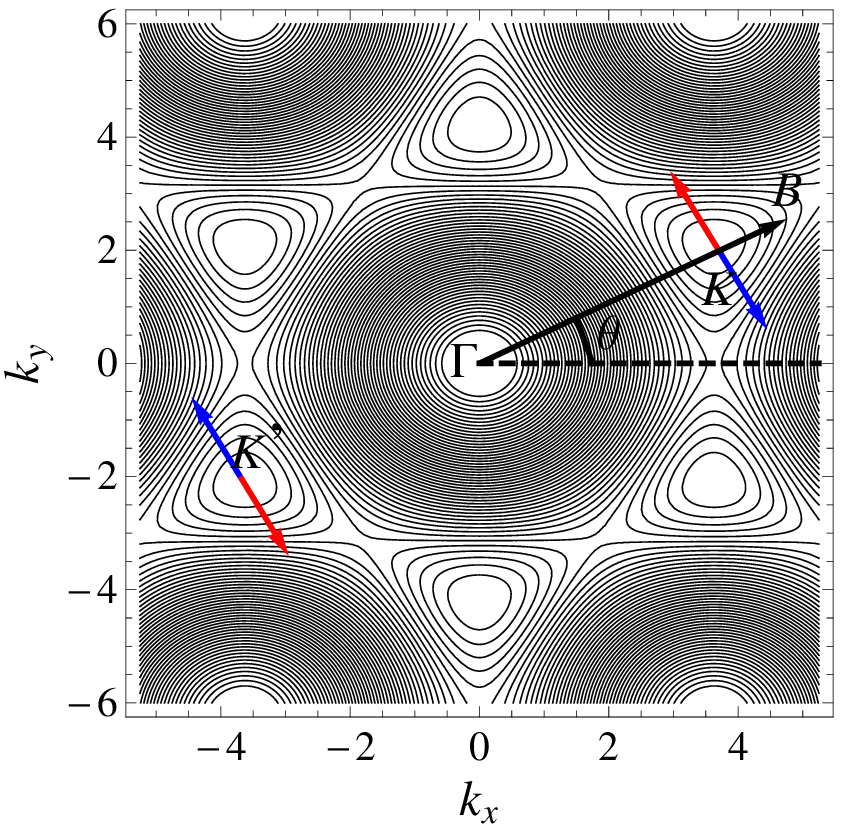}
\includegraphics[width=4.25cm,height=3.75cm]{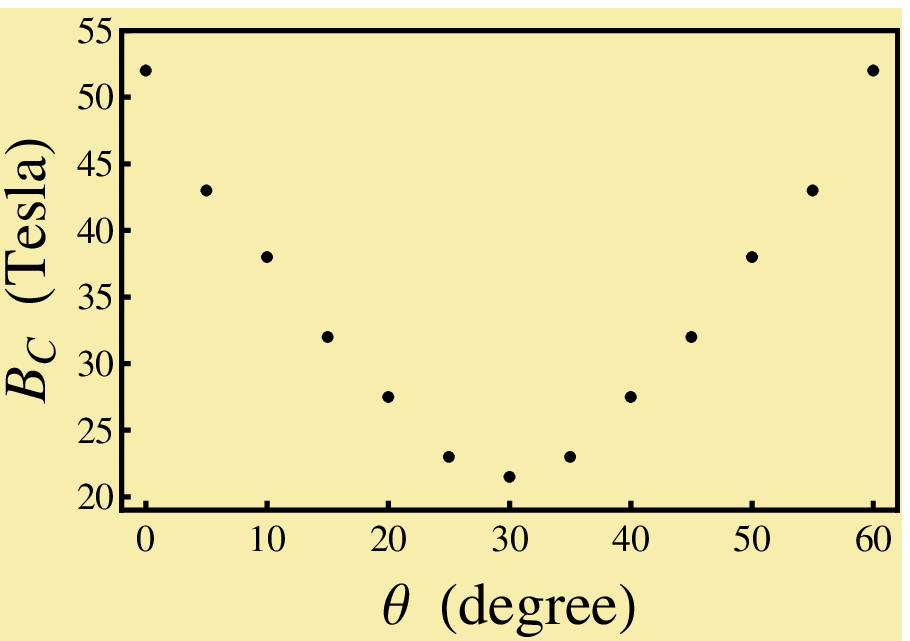}
\caption{(Color online)Left: Splitting of the Dirac points (red and blue arrows) due to in-plane field (B) (here $K'=-K$). $k_x,k_y$ are measured in unit $\hbar=a=1$. Right: A schematic variation of $B_{C}$ with $\theta$ (see left), with $v_F =10^6 m/s$, $t_\perp/t=10$, $v_F/v_2=30$, $d=3.5 \mathring{A}$.\cite{band-parameters-1,band-parameters-2} }
\label{Bcritical}
\end{figure}

To construct the low energy theory, we first neglect electron spin for simplicity, and define an 8-component spinor $\Psi^\top_B(\vec{k})$ $=\left(\Psi_+, \Psi_-\right)(\vec{k})$, with $\Psi^\top_\pm (\vec{k})=\big[ v_1( \vec{K} \pm \vec{k}_B + \vec{k})$, $v_2( \vec{K} \pm \vec{k}_B + \vec{k})$, $v_1(- \vec{K} \pm\vec{k}_B + \vec{k})$, $v_2(- \vec{K} \pm \vec{k}_B + \vec{k})\big]$, which account for the layer, valley and the $\pm \vec{k}_B$ (hereafter referred as \emph{flavor}) degrees of freedom. In this basis $H[B]$ takes the relativistic invariant form  $H_D= -v_{B} i \Gamma_0  \left( \Gamma_1 \; k_x + \Gamma_2 \; k_y \right)$, where $i\Gamma_0 \Gamma_1$ $=332 $, $i\Gamma_0 \Gamma_2$ $=301$, and $\Gamma_0=003$.\cite{compactnotation} An identical Hamiltonian also describes the low energy theory in twisted BLG, when the twisting is commensurate, with $v_B$ $\rightarrow$ $v_T$. The effective Fermi velocities near the new Dirac points
\begin{equation}
v_B=\frac{v^2_F B d}{t_\perp} \sim 1.2 \times 10^3 B(\mathrm{T}) \frac{m}{s}, \; v_T = |\vec{K}| \sin{\phi} \frac{v^2_F}{\tilde{t}_\perp},
\end{equation}
can be tuned by the in-plane magnetic field and twisting, respectively. Here $\tilde{t}_\perp \approx 0.4 t_\perp$\cite{castro-twisted-1, Li-fermivelocity}, and $\phi$ measures deviation from the AB-stacking. Respectively, in these two systems the dispersion is linear over the energy $v_B \Lambda_B \approx 10^{-5} B^2$(T) K, and $v_T \Lambda_T \approx 1.76 \times 10^{4} \sin^2\phi$ K, where $\Lambda_B \approx$ $5.6 \times 10^{-29} B$(T) kg m/s and $\Lambda_T \approx \hbar \sin{\phi}/a$ are the associated cut-offs for momentum. Therefore, the effect of twisting is much stronger than in-plane field. However, when the in-plane field and twisting are present together, one can tune Dirac dispersion continuously and we focus on this setup, with Fermi velocity $v_F$ and cutoff $\Lambda$ coming from the combined effects of twisting (we consider only commensurate ones) and in-plane field and continuously tunable, unless specifically noted otherwise. The imaginary-time Lagrangian is $L_0= \Psi^\dagger_B(\tau, \vec{k}) \left( \partial_\tau+ H_D \right) \Psi_B(\tau, \vec{k})$. The Dirac Hamiltonian $H_D$ commutes with the generator of translation $I_{Tr}= 3 3 0$.\cite{compactnotation} It also commutes with $I_{L}=001$, and $I_{K}=010$, when accompanied by the momentum axis inversion $k_x \rightarrow -k_x$. Respectively, these two operators exchange two layers and valleys. Moreover, $H_D$ is invariant under the exchange of the flavors, generated by $I_f$ $= 100$, after taking $\vec{k} \to -\vec{k}$. Additionally, $H_D$ is also invariant under emergent chiral $U_c(4)$ symmetry, generated by $\big\{x12,x22,x33,$ $x03,y11,y21,y30,y00 \big\}$, where $x=1,2$ and $y=0,3$.

If the repulsive interactions among the fermions are sufficiently strong, various ordered phases can be realized in this system. We assume there exist metallic gates nearby that render the Coulomb interaction to be short-ranged due to screening. In the long-wavelength limit the interacting Lagrangian for spinless fermions contains $27$ quartic terms of the form $\tilde{g}$ $(\Psi^\dagger_B M \Psi_B)^2$, out of which $18$ are independent.\cite{supplementary} Next we wish to capture the leading instabilities in this system in the large-$N$ scheme, in which all the couplings are independent. After integrating out the fast Fourier modes with the Matsubara frequencies $-\infty$ $<\omega<$ $\infty$ and $\Lambda/b<|\vec{k}|< \Lambda$, where $b>1$, we obtain the flow equations of dimensionless couplings $g=4 \tilde{g} \Lambda/(v_F \pi)$
\begin{equation}\label{betafunction}
\beta_g =\frac{d g}{d \; \log{b}}= -g - C_M \: g^2 + {\cal O} (\frac{1}{N}).
\end{equation}
The coefficient $C_M$ $=(1)0$, if $M$ (anti-)commutes with $H_D$, whereas $C_M=1/2$ if $M$ commutes with either $i \Gamma_0 \Gamma_1$ or $i \Gamma_0 \Gamma_2$. Therefore, the leading instabilities in this system take place towards the formation of the fully gapped or massive\cite{mass} phases at $T=0$, which minimizes the energy of the filled Dirac-Fermi sea\cite{graphene-interaction-1}. The linear term in the $\beta$-functions indicates that interactions need to be sizable to place the system in any ordered phase. Depending on wether $g>$ or $<1$ at the scale $\Lambda$, which is determined by the marginally relevant flow of pristine BLG at momentum scale beyond $\Lambda$\cite{vafek-kun,fanzhang}, the Dirac fermions find themselves in ordered or semi-metallic phase. Since $\Lambda$ is a {\em tunable} parameter here, BLG subject to twisting and in-plane fields yields unique opportunities to observe various relativistic quantum critical phenomena, described by the Gross-Neveu-Yukawa (GNY) theory, about which in a moment. We note our estimation from the large-N analysis is expected to hold even for the realistic situation when $N$ (number of 8-component Dirac fermions)$=1$, since the actual expansion parameter in our theory is $1/8 N$.\cite{1oNcorrection}

\begin{center}
\begin{tabular}{|l|l|l|l|l|l|}
\hline
$\vec{M}$ & \hspace{0.65cm}$I_L$ & \hspace{0.65cm}$I_K$ & \hspace{0.6cm}$I_{Tr}$ & \hspace{0.6cm}$I_f$ & \hspace{0.5cm}$I_T$ \\\hline \hline
$\vec{A}$ & (-,+,+,-) & (+,+,+,+) & (+,-,-,+) & (-,-,+,+) & (-,+,+,+)   \\\hline
$\vec{B}$ & (-,+,+,-) & (-,-,-,-) & (+,-,-,+) & (-,-,+,+) & (+,-,-,-)    \\\hline
$\vec{C}$ & (-,+,+,-) & (+,+,+,+) & (-,+,+,-) & (-,-,+,+) & (+,+,+,-)     \\\hline
$\vec{D}$ & (-,+,+,-) & (-,-,-,-) & (-,+,+,-) & (-,-,+,+) & (+,+,+,-)      \\\hline
\end{tabular} \vspace{0.3cm}\\
\end{center}
Table I: Transformation of various masses under the discrete symmetries, which leave the free Hamiltonian invariant. $+$, $-$ respectively stand for even and odd. Here, $I_t=(110)$K is the time-reversal operator, where K is the complex conjugation.

The spinless flavored Dirac fermions can condense into a plethora of fully gapped phases. Altogether, there are $16$-different ways to spontaneously develop mass gaps for the spinless Dirac fermions\cite{mass}. They can be arranged into $4$ categories, and each of them contain $4$ masses. They are $\vec{A}$ $=\left\{303, 200, 100, 003 \right\}$, $\vec{B}$ $=\left\{333, 230, 130, 033 \right\}$,  $\vec{C}$ $=\left\{312,  211, 111, 012\right\}$, $\vec{D}$ $=\left\{322, 221, 121, 022\right\}$\cite{compactnotation}. The transformations of these masses under various discrete symmetries are shown in Table I. $\vec{A}$, $\vec{B}$ do not mix two valleys, but are respectively even and odd under $\vec{K} \leftrightarrow -\vec{K}$, whereas $\vec{C}$ and $\vec{D}$ represents symmetric and antisymmetric mixing of two valleys, respectively. All the \emph{four} masses of the pristine BLG\cite{vafek-RG} are flavor insensitive: The layer polarized (LP) state, corresponding to an imbalance of carriers densities between two layers, is represented by $A_4$; $B_4$ represents the quantum anomalous Hall insulators, which in lattice is realized as intra-layer circulating currents, orienting in the same direction in two layers. Two translational and time-reversal symmetry breaking Kekule currents are represented by $C_4$ and $D_4$. Eight out of the above sixteen masses, $X_j$s, where $X=A,B, C, D$; $j=2,3$, mix the flavors, and correspond to periodic orders with periodicity $2 \Lambda$. The remaining four masses $X_1$s where $X=A,B,C,D$, although they do not mix the flavors, are odd under its exchange.

We now restore the spin degrees of freedom. Each of the masses can now be realized in both spin-singlet and spin-triplet channels, yielding all together $2 \times 16=32$ possible insulating orders that can fully gap out all the Dirac points. The spin-triplet insulators, besides the discrete symmetries, also break the spin $SU(2)$ rotational symmetry spontaneously (if there is no Zeeman splitting say in twisted BLG with no in-plane field), and the ordered phases are accompanied by $2$ massless Goldstone modes. Let us construct a 16-component spinor $\Psi_{s}=\left( \Psi_\uparrow, \Psi_\downarrow \right)^\top$, where $\uparrow, \downarrow$ are the electron's spin projections. This representation is invariant under the electron's spin rotations, which are generated by $\vec{S}=\vec{s}\otimes (000)$.\cite{compactnotation} Both $\Psi_\uparrow$ and $\Psi_\downarrow$ take the form $\Psi_B(\vec{k})$. The Dirac Hamiltonian for spinful fermions is $H=s_0 \otimes H_D$, where $s_0$ is a two dimensional identity matrix operating on the spin index, and  $[H,\vec{S} ]=0$. The Zeeman coupling, when present, reads as $H_Z= \Delta_Z \; \left( s_3 \otimes I_8 \right)$, where $\Delta_Z =g B(\vec{x})$ $\sim B$ (T) K, with $g \approx 2$ for electrons in BLG. We note $\Delta_Z \gg v_B \Lambda_B$ for any accessible magnetic field, but $\Delta_Z \ll v_T \Lambda_T$ even for moderate twisting, say $\phi \sim 3^0-4^0$.

While small compared to twisting, Zeeman coupling gives rise to particle- and hole-like Fermi surfaces for opposite spin projections near each Dirac points, and consequently a BCS instability in the particle-hole triplet channel takes place even at weak interactions\cite{keldysh-kopaev}. However the critical temperature of such ordering can be extremely small.\cite{aleiner-tsvelik-kharzeev} Perhaps more importantly, the Zeeman coupling restricts the spin degrees of freedom of any triplet OP, ${\cal M}=\vec{\Delta} \cdot \vec{s} \otimes X$, where $\{ X, H_D \}=0$, within the easy plane, perpendicular to the applied magnetic field. The effective single-particle Hamiltonian then reads as $H_{SP}=H+{\cal M}+H_Z$, and its energy eigenvalues are $\pm E_\sigma$, where for $\sigma=\pm$
\begin{equation}
E_\sigma= \bigg\{ \left[ \sqrt{v^2_{F} k^2+\Delta^2_3} + \sigma \; \Delta_Z \right]^2 + \Delta^2_1 +  \Delta^2_2 \bigg\}^{1/2}.
\end{equation}
Therefore, the spectrum is maximally gapped with $\Delta_3=0$, or when the triplet OP is restricted within the easy plane. Consequently, one of the Goldstone modes becomes massive, and its mass is $\sim \Delta_Z$. Hence, the Zeeman coupling reduces the symmetry of any triplet ordering to $O(2)$, restoring the possibility of Kosterlitz-Thouless transitions at finite temperatures.\cite{kosterlitz-thouless} Recent experiments\cite{BLG-experiment4, BLG-experiment5} suggest that a layer anti-ferromagnet (LAF) state can be found in BLG. The LAF OP is $\langle \Psi^\dagger_s \left(\vec{s} \otimes A_4 \right)\Psi_s \rangle$. Therefore, subject to an in-plane magnetic field, spin of the LAF order gets projected onto the easy plane, known as the \emph{canted anti-ferromagnet}, which has also been considered in the quantum Hall regime of insulating BLG\cite{roy-QHE-BLG}, and single-layer graphene\cite{herbut-so3}. In twisted BLG, beyond the super-lattice AB-stacking goes through an AA one and evolves into BA stacking\cite{sanjose-guinea-gonzalez}. Therefore, the layer magnetization changes its sign beyond the super-lattice; however, the LAF order remains unchanged. A similar situation also arises when the system develops a LP state.

The quantum phase transition towards the formation of the LAF state, driven solely by the twisting, is described by an $O(3)$ GNY theory. A similar theory can also describes the quantum criticality near the anti-ferromagnet ordering in monolayer graphene, where the number of 4-component Dirac fermions is two, which in our system is four.\cite{sO3-AF} The Fermi velocity across the transition remain non-critical since $z=1$ in our problem\cite{graphene-interaction-1}. The effective action reads as $S=\int d^dx L$, where $L= L_f + L_b + L_{bf}$, with $L_f = \bar{\Psi}_s s_0 \otimes \Gamma_\mu \partial_\mu \Psi_s$, with $\bar{\Psi}_s=\Psi^\dagger_s s_0 \otimes \Gamma_0$ and
\begin{equation}\label{O3GNY}
L_b = |\partial_\mu \vec{\Phi}|^2 + m^2_t |\vec{\Phi}|^2 + \frac{\lambda_t}{2} |\vec{\Phi}|^4, L_{bf}= g_t \vec{\Phi} \cdot \bar{\Psi}_s \vec{s} \otimes I_8 \Psi_s.
\end{equation}
$\vec{\Phi}$ is a three-component scalar field, and $g_t$ is the Yukawa coupling. The coupling constants $\lambda_t$, $g_t$ are dimensionless in $d=3+1$, and $m^2$ is the $T=0$ tuning parameter. Upon promoting the theory to the upper-critical dimension ($d=4$), we can perform a controlled $\epsilon(=d-4)$-expansion. In the absence of the Yukawa coupling the transition is described by the Wilson-Fisher fixed point $(\lambda_t,g^2_t)=\left(6/11,0 \right)\epsilon$.\cite{supplementary} However, this fixed point is unstable against the Yukawa coupling \cite{sO3-AF,supercondcriticality}, and the critical behavior of the GNY theory is described by a new fixed point $(\lambda_t,g^2_t)=\left(0.61942,0.11 \right)\epsilon$. Near the $O(3)$ LAF transition the correlation length exponent is $\nu=\frac{1}{2}+ 0.531268 \epsilon$. The same theory can describe the critical behavior near any triplet ordering in twisted BLG. Due to the non-trivial Yukawa coupling at the critical point, both the bosonic and the fermionic fields acquire nontrivial anomalous dimensions, respectively read as $\eta_b=\frac{8}{9} \epsilon$, $\eta_f=\frac{\epsilon}{6}$.\cite{supplementary} The bi-critical fixed point in the GNY theory lies in the unphysical regime $\lambda_t <0$. As one approaches the critical point from the semi-metallic side, the residue of the quasi-particle pole vanishes as $m_t^{z \nu \eta_f}$.

We now consider driving the Dirac semi-metal across the LAF transition by tuning a parallel magnetic field, in the presence of twisting. In that situation the universal behavior in the vicinity of the LAF transition will be governed by an $O(2)$ GNY theory, due to the Zeeman coupling. However, such quantum critical behavior can only be probed at a temperature $T>\Delta_Z$. The effective theory is similar to the one in Eq.~(\ref{O3GNY}), with $\vec{\Phi}$ as a two-component bosonic field, and $\vec{s} \rightarrow \vec{s}_\perp=(s_1,s_2)$. A similar $O(2)$ GNY theory also describes the quantum superconducting transition of the Dirac fermions in graphene and on the surface of topological insulators\cite{supercondcriticality}, and in-plane field driven transition to any other triplet ordering in twisted BLG. Various critical exponents near this transition are different from the previous ones and read as $\nu=\frac{1}{2}+ \frac{3}{10} \epsilon$, $\eta_b=\frac{4}{5} \epsilon$, and $\eta_f=\frac{\epsilon}{10}$.\cite{supercondcriticality}

As a consequence of the restoration of relativistic invariance at the GNY critical point, i.e. $z=1$, the bosonic velocity approaches the the fermionic velocity ($v_F$) (non-universal), and consequently the ratio of the specific heats inside the semi-matallic and insulating side within the critical region also approaches universal value 
\begin{equation}
\frac{C_{SM}}{C_{Ins}}=\frac{N_f}{N_G} \; \left(1-2^{-d} \right),
\end{equation} 
where $N_f=16$ is the number of gapless fermionic modes, and $N_G$ is the number of massless Goldstone modes in the broken-symmetry phase, which is therefore respectively $2$ and $1$ in the twisting and the Zeeman-driven LAF phase.

In the same framework, we can also address the quantum critical behavior of a $Z_2$ symmetry breaking transition towards the LP ordering in twisted BLG with \cite{supplementary}
\begin{equation}
L_b = |\partial_\mu \Phi|^2 + m^2_s |\Phi|^2 + \frac{\lambda_s}{2} |\Phi|^4, L_{b-f}= g_s \Phi \bar{\Psi}_s \Psi_s.
\end{equation}
The transition to a $Z_2$ symmetry breaking ordering is governed by the critical point $\left( \lambda_s, g^2_s\right)=(0.5914, 0.091) \epsilon$. The critical exponents near this critical point are $\nu=\frac{1}{2}+ 0.25574 \epsilon$, $\eta_b= \frac{8}{11}\epsilon$, and $\eta_f=\frac{\epsilon}{22}$\cite{supplementary}. At $T=0$, however, there is a first order phase transition from the LP to the $O(2)$ LAF state, which also carries a finite ferromagnetic moment, if one applies a weak in-plane field.

Finally we propose a possible way to realize some of the new masses, which lack any analogy in the pristine BLG. In the vicinity of the neutrality point, we believe that the leading instabilities will likely be similar to those in regular BLG, when it is slightly twisted and placed in weak parallel fields. On the other hand, upon tilting the magnetic field out of the BLG plane, one can develop a set of Landau levels (LLs) at energies $\pm v_{F} \sqrt{2 n B_\perp}$, where $n=0,1,\cdots$ and $B_\perp$ is the field's perpendicular component. Relativistic LLs can also be developed by placing the twisted BLG in perpendicular magnetic fields. The twofold \emph{orbital degeneracy} of the zeroth LL in pristine BLG\cite{falko-Mccann}, translates into the \emph{flavor degeneracy} when BLG is twisted\cite{castro-twisted-1} or when twisting and tilted fields are present simultaneously, which however, can not be lifted by any of the masses in pristine BLG, e.g $A_4$, $C_1$, $D_1$\cite{roy-classification}, since they are flavor independent. If we consider spinless fermions for simplicity, and also neglect flavor and valley mixing, the flavor degeneracy of the zeroth LL can be lifted by $A_1$ or $B_1$ masses. Therefore, by placing the chemical potential close to the first excited state within the zeroth LL, additional incompressible Hall states at fillings $f=\pm 1$ can be developed by these new masses that we propose here.\cite{barlas} The single-particle gap of the $f=0,\pm 1$ Hall states should scale linearly with the field, if the interaction is sufficiently weak. Scaling then reverts to a sublinear one for moderate interaction strength; and a perfect $\sqrt{B}$ scaling emerges at zero-field criticality\cite{roy-scaling}. Interestingly, different scaling regimes can be accessed in this system by tuning the twisting and/or in-plane field, which in turn controls the effective interaction strength. A detailed analysis of the quantum Hall physics for spinful fermions is quite rich, but left for future investigation.

B. R. was supported at National High Magnetic Field Laboratory by NSF Cooperative Agreement No.DMR-0654118, the State of Florida, and the U. S. Department of Energy. K.Y. is supported by NSF Grant No. DMR-1004545. We thank Igor F. Herbut and Oskar Vafek for valuable discussions.

\pagebreak

\onecolumngrid

\section*{Supplementary Material for EPAPS}

\section{derivation of the low energy Hamiltonian in BLG, subject to in-plane magnetic field}

In this section we present the detailed derivation of the low energy Hamiltonian in BLG, subject to an in-plane magnetic field $\vec{B}=B \left( \cos \theta, \sin \theta, 0\right)$. In the absence of the in-plane magnetic field, the four band tight binding Hamiltonian in the vicinity of two inequivalent valleys, suitably chosen here at $\pm \vec{K}$, where $\vec{K}=\left(1,1/\sqrt{3}\right) 2\pi /(\sqrt{3}a)$ and $a \approx 2.5 \mathring{A}$ is the lattice spacing in each layer of BLG, takes the form $H_+ \oplus H_-$, where
\begin{equation}
H_\pm = \left( \begin{array}{c c c c}
0 & t_\perp & v_2 \left( k_x \mp i k_y \right) & v_F \left( k_x \pm i k_y \right) \\
t_\perp & 0 & v_F \left( k_x \mp i k_y \right) & v_2 \left( k_x \pm i k_y \right) \\
v_2 \left( k_x \pm i k_y \right) & v_F \left( k_x \pm i k_y \right) & 0 & 0 \\
v_F \left( k_x \mp i k_y \right)  & v_2 \left( k_x \mp i k_y \right) & 0 & 0
\end{array}
\right).
\end{equation}
Here the four component spinor is  defined as $\Psi^{\top}(\vec{k})=\left( u_1, u_2,v_2,v_1\right)(\vec{k})$, where $u_j(v_j)$ is the fermionic annihilation operator on $A(B)$  sub-lattice in $j^{th}$ layer, with $j=1,2$. The energy spectrum of the above non-interacting Hamiltonian is composed of four bands. Two out of which are fully gapped, with band gap $\sim 2 t_\perp$. These two bads are predominantly localized on A-sublattices on two layers. Therefore, to obtain the effective low energy Hamiltonian we will integrate out the high energy sites. We, however, perform this exercise in the presence of an in-plane magnetic field. The orbital effect of the in-plane magnetic field can be captured via the minimal coupling as follows
\begin{eqnarray}
k_x \to k_x + \frac{B \cdot d}{2} \sin \theta, \:   k_y \to  k_y - \frac{B \cdot d}{2} \cos \theta \; : \: \mbox{in the intra-layer hopping integral in the bottom layer,} \nonumber \\
k_x \to k_x - \frac{B \cdot d}{2} \sin \theta, \:  v_F k_y \to v_F k_y + \frac{B \cdot d}{2} \cos \theta \; : \: \mbox{in the intra-layer hopping integral in the top layer.} \nonumber
\end{eqnarray}
For our convenience, we define two spinors as $\psi^\top_A(\vec{k})= \left(u_1,u_2 \right)(\vec{k})$ and $\psi^\top_B(\vec{k})= \left(v_1,v_2 \right)(\vec{k})$, and the effective low energy Hamiltonian will be written in the basis of $\psi_B$ spinor. Otherwise, in the vicinity of the $ \pm \vec{K}$ valleys, and to the quadratic order in momentum it reads as $H^{eff}_\pm$, where
\begin{equation}
H^{eff}_\pm= H_{BA,\pm} \: G_{AA} (0) \: H_{AB,\pm} + H_{BB,\pm}.
\end{equation}
Various two dimensional matrices in the above equation are
\begin{equation}
H_{AB,\pm}=\left( \begin{array} {c c}
v_F \left[ (k_x + \frac{B \cdot d}{2} \sin \theta) \pm i (k_y - \frac{B \cdot d}{2} \cos \theta) \right] & 0 \\
0 & v_F \left[ (k_x - \frac{B \cdot d}{2} \sin \theta) \mp i (k_y + \frac{B \cdot d}{2} \cos \theta) \right]
\end{array}
\right),
\end{equation}
\begin{equation}
H_{BA,\pm}=\left( \begin{array} {c c}
v_F \left[ (k_x + \frac{B \cdot d}{2} \sin \theta) \mp i (k_y - \frac{B \cdot d}{2} \cos \theta) \right] & 0 \\
0 & v_F \left[ (k_x - \frac{B \cdot d}{2} \sin \theta) \pm i (k_y + \frac{B \cdot d}{2} \cos \theta) \right]
\end{array}
\right),
\end{equation}
and
\begin{equation}
G_{AA} (i \omega)= \frac{i \omega I_2 + \sigma_1 t_\perp}{\omega^2 + t^2_\perp}, \: H_{BB,\pm}= v_2 \left( \pm \sigma_2 k_x - \sigma_1 k_y \right).
\end{equation}
After performing the matrix multiplications, and defining a four component spinor as $\Psi(\vec{k})=\left( v_1(\vec{K}+\vec{k}), v_2(\vec{K}+\vec{k}), v_1(-\vec{K}+\vec{k}), v_2(-\vec{K}+\vec{k}) \right)$, we obtain the following effective low energy Hamiltonian
 \begin{equation}
H[B] =  v_2 i \gamma_0 \gamma_1 k_x + v_2 i\gamma_0 \gamma_2 k_y + \gamma_2 \; \bigg( \frac{k^2_x-k^2_y}{2 m}  \nonumber \\
+  \frac{k^2_{Bx}-k^2_{By}}{2 m} \bigg) - \gamma_1 \left( \frac{ 2 k_x k_y}{2m} + \frac{ 2 k_{Bx} k_{By}}{2m}\right),
\end{equation}
as shown in Eq.~(\ref{effectiveMag}) in the main text.

\section{Generalized interacting Lagrangian for spinless fermions}

In the absence of the in-plane/parallel magnetic fields or twisting, there is no \emph{flavor} degrees of freedom. Then for the spinless fermions there are all together $9$ quartic terms that defines the interacting Lagrangian in pristine BLG, which reads as \cite{vafek-prb}
\begin{eqnarray}
L^{BL}_{int}&=&g_{A_{1}} \left( \Psi^\dagger \Psi \right)^2 + g_{A_{2}} \left\{ \left( \Psi^\dagger \gamma_1 \Psi \right)^2 +\left( \Psi^\dagger \gamma_2 \Psi \right)^2 \right\}+ g_{B_{1}} \left\{ \left( \Psi^\dagger i \gamma_0 \gamma_1 \Psi \right)^2 +\left( \Psi^\dagger i \gamma_0 \gamma_2 \Psi \right)^2 \right\}+g_{A_{1}} \left( \Psi^\dagger  i \gamma_3 \gamma_5 \Psi \right)^2 \nonumber \\
&+& g_{C_{1}} \left( \Psi^\dagger \gamma_0 \Psi \right)^2 + g_{D_{2}} \left( \Psi^\dagger i \gamma_1 \gamma_2 \Psi \right)^2 + g_{\alpha} \left\{ \left( \Psi^\dagger i \gamma_0 \gamma_3 \Psi \right)^2 +\left( \Psi^\dagger i \gamma_0 \gamma_5 \Psi \right)^2 \right\} + g_{\gamma} \left\{ \left( \Psi^\dagger \gamma_3 \Psi \right)^2 +\left( \Psi^\dagger \gamma_5 \Psi \right)^2 \right\} \nonumber \\
&+& g_{\alpha} \left\{ \left( \Psi^\dagger i \gamma_1 \gamma_3 \Psi \right)^2 +\left( \Psi^\dagger i \gamma_1 \gamma_5 \Psi \right)^2 + \left( \Psi^\dagger i \gamma_2 \gamma_3 \Psi \right)^2 +\left( \Psi^\dagger i \gamma_2 \gamma_5 \Psi \right)^2\right\}.
\end{eqnarray}
To remind ourselves, the four-component spinor reads as $\Psi^\top(\vec{k})=\big[ v_1(\vec{K}+\vec{k})$, $v_2(\vec{K}+\vec{k})$, $v_1(-\vec{K}+\vec{k})$, $v_2(-\vec{K}+\vec{k}) \big]$, suitable to capture the low energy behavior in regular BLG. The representation of the $\gamma$ matrices can be obtained from the main body of the paper. However, not all the \emph{nine} coupling constants are linearly independent, and there exist a set of algebraic constraints, \emph{Fierz identity}, which allow us to write each of the quartic term as linear combination of the rests. It can further be shown that there are only \emph{four} linearly independent coupling constant\cite{HJR}. Upon applying an in-plane field, the parabolic touching of the bands splits into two Dirac cones (we here neglect the effect of trigonal warping, as we have shown that for sizable in-plane field, its effect can be neglected.), and we introduce the \emph{flavor} degrees of freedom. If we assume that the low energy theory is flavor independent, then the role of the flavor degrees of freedom is identical to that of the spin degrees of freedom in monolayer graphene. And consequently each of the four-fermion terms can be realized in \emph{flavor-singlet}, and \emph{flavor-triplet} channel. One can obtain the interacting Lagrangian by replacing
\begin{equation}
g_X \left( \Psi^\dagger M \; \Psi \right)^2 \to g_{X,s} \left( \Psi^\dagger_B \; \sigma_0 \otimes M \; \Psi_B \right)^2 + g_{X,t} \left( \Psi^\dagger_B \; \vec{\sigma} \otimes M \; \Psi_B \right)^2,
\end{equation}
where $\left( \sigma_0,\vec{\sigma} \right)$ are standard 2-dimensional unit and Pauli matrices, operating on the flavor index. Therefore, in BLG, placed in a magnetic field, the interacting Lagrangian is described by $18$ coupling constants. However, only \emph{nine} out of 18 coupling constants are linearly independent, and one can choose only the $g_{X,s}$'s as the independent couplings\cite{vafek-prb}. One should, however, notice a subtle difference between the spin degrees of freedom in monolayer graphene and the flavor one in BLG. In monolayer graphene the non-interacting Hamiltonian is \emph{spin-independent}, whereas in BLG, placed in a parallel magnetic field, the non-interacting Hamiltonian changes sign upon interchanging the flavors. As a result, each flavor-triplet interactions break into two components, \emph{flavor-easy-plane}, and \emph{flavor-easy-axis}, as follows
\begin{equation}
 g_{X,t} \left( \Psi^\dagger_B \; \vec{\sigma} \otimes M \; \Psi_B \right)^2 \to g^\perp_{X,t} \left( \Psi^\dagger_B \; \vec{\sigma}_\perp \otimes M \; \Psi_B \right)^2 + g^\parallel_{X,t} \left( \Psi^\dagger_B \; \sigma_\parallel \otimes M \; \Psi_B \right)^2,
\end{equation}
where $\vec{\sigma}_\perp=\left( \sigma_1,\sigma_2\right)$, and $\sigma_\parallel=\sigma_3$. Hence all together there are 27 coupling constants. Only 18 out of them are linearly independent, which can for example be chosen to be $g_{X,s}$s and $g^\perp_{X,t}$. However, in the large-N limit there is no linear relationship between these coupling constants, and all 27 couplings are independent. We will perform our renormalization group analysis in the large-N framework.

\section{Renormalization group analysis of Gross-Neveu-Yukawa theory}

We now present the detail of the renormalization group calculation for the $Z_2$ and $O(3)$ phase transitions, in the framework of GNY theory\cite{Zinn-Justin, herbut-juricic-vafek}. In the former situation the renormalized action reads as
\begin{equation}
S= \int d^dx \left\{ Z_{\Phi,s} \left(\partial_\mu \Phi \right)^2  +m^2_{s,0} \Phi^2 +\frac{\lambda_{s,0}}{2} \Phi^4  + Z_{\Psi,s} \left(\bar{\Psi}_s s_0 \otimes \Gamma_\mu \partial_\mu \Psi + g_{s,0} Z^{1/2}_{\Phi,s} \Phi \bar{\Psi} \Psi\right)	\right\}.
\end{equation}
The quantities with subscript $``0"$ correspond to the bare values. Otherwise, the renomalization conditions read as
\begin{eqnarray}
Z_{\Psi,s}= 1- \frac{1}{2} \; g^2_s \; \frac{m^{-\epsilon}_s}{\epsilon} + {\cal O}(1); \: \:
Z_{\Phi,s}=1-2 \; g^2_s \; \frac{N}{\epsilon}+ {\cal O} (1); \: \:
m^{-\epsilon/2}_s g_{s,0} Z^{1/2}_{\Phi,s} Z_{\Psi,s} - g^3_s m^{-\epsilon}_s \frac{1}{\epsilon}=g_s
\end{eqnarray}
\begin{equation}
m^{-\epsilon}_s \lambda_{s,0} Z^2_{\Phi,s} + 24 \; g^4_s \; \frac{N }{\epsilon}-\frac{3}{2} \lambda^2_s \; \frac{1}{\epsilon}=\lambda_s; \: \:
Z_{\Phi,s} m^2_{s,0} -\frac{\lambda_s}{2} \; m^2_{s,0} \frac{1}{\epsilon} = m^2_s,
\end{equation}
where $N$ is the number of four component Dirac fermions. From these conditions, we obtain the flow equations or the $\beta$-functions of the various quantities as follow
\begin{eqnarray}
\beta_{g^2_s}=\epsilon \; g^2_s - (2 N +3) g^4_s; \:\:
\beta_{\lambda_s}=\epsilon \lambda_s -\frac{3}{2} \lambda^2_s - 4 N \lambda_s g^2_s + 24 N g^4_s; \: \:
\beta_{m^2_s}=2 m^2_s - \frac{1}{2} \; \lambda_s m^2_s - 2 N g^2_s m^2_s.
\end{eqnarray}
The above flow equations discerns the following fixed points in the critical plane, $m^2_s=0$,
\begin{eqnarray}
\left( \lambda_s, g^2_s\right) = \left(0,0 \right) \; \left( \mathrm{Gaussian}\right); \: \:
\left( \lambda_s, g^2_s\right) = \left(\frac{\left(3-2 N\right) - \sqrt{9+ 4 N (33+N)}}{9+6 N}\; \epsilon,\frac{\epsilon}{2 N+3} \right) \; \left( \mathrm{bi-critical}\right) \nonumber
\end{eqnarray}
\begin{eqnarray}
\left( \lambda_s, g^2_s\right) = \left(\frac{2}{3}\; \epsilon,0 \right) \; \left( \mathrm{Wilson-Fisher}\right); \: \:
\left( \lambda_s, g^2_s\right) &=& \left(\frac{ \left(3-2 N\right) + \sqrt{9+ 4 N (33+N)}}{9+6 N}\; \epsilon,\frac{\epsilon}{2 N+3} \right) \; \left( \mathrm{critical}\right) \nonumber \\
&=& \left( \lambda^\ast_s,(g^2_s)^\ast\right); \nonumber
\end{eqnarray}
From here we can find various critical exponents as follows
\begin{equation}
\nu= \frac{1}{2}+ \frac{1}{2} \left[ \frac{1}{4} \lambda^\ast_s + N (g^2_s)^\ast \right]; \: \: \:
\eta_\Psi = \frac{1}{2} (g^\ast_s)^2=\frac{1}{2} \; \left( \frac{\epsilon}{3+2 N}\right); \: \:
\eta_\Phi = 2 N \; (g^\ast_s)^2= 2N \; \left( \frac{\epsilon}{3+ 2 N}\right).
\end{equation}
Results, quoted in the main text, are obtained by setting $N=4$.

Pursuing a similar approach one can obtain the critical behavior in the vicinity of the  $O(3)$ LAF transition, which is pertinent for the twisting driven quantum phase transition. The renormalized action in this situation reads as
\begin{equation}
S= \int d^dx \left\{ Z_{\Phi,t} \left(\partial_\mu \vec{\Phi} \right)^2  +m^2_{t,0} \Phi^2 +\frac{\lambda_{t,0}}{2} (\vec{\Phi}\cdot \vec{\Phi})^2  + Z_{\Psi,t} \left(\bar{\Psi}_t s_0 \otimes \Gamma_\mu \partial_\mu \Psi + g_{t,0} Z^{1/2}_{\Phi,t} \; \vec{\Phi} \cdot \bar{\Psi} \left( \vec{s} \otimes I_8 \right) \Psi\right) \right\}.
\end{equation}
The renormalization conditions are
\begin{eqnarray}
Z_{\Psi,t}= 1- \frac{3}{2} \; g^2_t \; \frac{m^{-\epsilon}_t}{\epsilon} + {\cal O}(1); \: \:
Z_{\Phi,t}=1-2 \; g^2_t \; \frac{N}{\epsilon}+ {\cal O} (1); \: \:
m^{-\epsilon/2}_t g_{t,0} Z^{1/2}_{\Phi,t} Z_{\Psi,t} - g^3_t m^{-\epsilon}_t \frac{1}{\epsilon}=g_t,
\end{eqnarray}
\begin{equation}
m^{-\epsilon}_t \; \lambda_{t,0} \; Z^2_{\Phi,t} + 24 \; g^4_t \; \frac{N}{\epsilon}-\frac{11}{6} \lambda^2_t \; \frac{1}{\epsilon}=\lambda_t; \: \:
Z_{\Phi,t} \; m^2_{t,0} -\frac{5}{6} \; \lambda_t \; m^2_{t,0} \; \frac{1}{\epsilon} = m^2_t,
\end{equation}
The infra-red renormalization group flow equations are
\begin{eqnarray}
\beta_{g^2_t}=\epsilon \; g^2_t - (2 N +1) g^4_t; \:\:
\beta_{\lambda_t}=\epsilon \lambda_t -\frac{11}{6} \lambda^2_t - 4 N \lambda_t g^2_t + 24 N g^4_t; \: \:
\beta_{m^2_t}=2 m^2_t - \frac{5}{6} \; \lambda_t m^2_t - 2 N g^2_t m^2_t.
\end{eqnarray}
From these flow equations, we obtain the following fixed points in the critical plane $m^2_t=0$
\begin{eqnarray}
\left( \lambda_t, g^2_t\right) = \left(0,0 \right) \; \left( \mathrm{Gaussian}\right); \: \:
\left( \lambda_t, g^2_t\right) = \left(\frac{3 \left[ \left(1-2 N\right) - \sqrt{1+ 4 N (43+N)} \right]}{11 \; \left(1+2 N \right)}\; \epsilon,\frac{\epsilon}{2 N+1} \right) \; \left( \mathrm{bi-critical}\right) \nonumber
\end{eqnarray}
\begin{eqnarray}
\left( \lambda_t, g^2_t\right) = \left(\frac{6}{11}\; \epsilon,0 \right) \; \left( \mathrm{Wilson-Fisher}\right); \: \:
\left( \lambda_t, g^2_t\right) &=& \left(\frac{3 \left[ \left(1-2 N\right) + \sqrt{1+ 4 N (43+N)} \right]}{11 \; \left(1+2 N \right)} \epsilon,\frac{\epsilon}{2 N+1} \right) \; \left( \mathrm{critical}\right) \nonumber \\
&=& \left( \lambda^\ast_t,(g^2_t)^\ast\right); \nonumber
\end{eqnarray}
One can read off various critical exponents from the above flow equations, which are
\begin{equation}
\nu= \frac{1}{2}+ \frac{1}{2} \left[ \frac{5}{12} \lambda^\ast_t + N (g^2_t)^\ast \right]; \: \: \:
\eta_\Psi = \frac{3}{2} (g^\ast_t)^2=\frac{3}{2} \; \left( \frac{\epsilon}{1+2 N}\right); \: \:
\eta_\Phi = 2 N \; (g^\ast_t)^2= 2N \; \left( \frac{\epsilon}{1+ 2 N}\right).
\end{equation}
In the main part of the paper, results are presented after substituting $N=4$. It also worth pointing out that after substituting $N=2$, which corresponds to 2 flavors of 4-component Dirac fermions, describing the situation in monolayer graphene, we recover the critical exponents for the transitions to charge-density-wave and anti-ferromagnet ordering reported in Ref.~\onlinecite{herbut-juricic-vafek}.


\begin{thebibliography}{99}
\bibitem{twistdefinition} When two layers of BLG are neither perfectly Bernel (AB) or AA stacked, is referred here as twisted BLG.
\bibitem{AsitesBLG} Sites on A-sublattice are separated by $2 t_\perp \approx 200-300$ meV, and are of no dynamical importance at low energy.
\end{thebibliography}
\end{document}